**Techniques for supercharging academic writing with generative AI**
Generalist large language models can elevate the quality and efficiency of academic writing.

Zhicheng Lin
zhichenglin@gmail.com
Department of Psychology, University of Science and Technology of China.

To many researchers, academic writing evokes a Sisyphean ordeal: it robs precious time and mental bandwidth that may be better spent doing actual science. Franz Kafka expressed it eloquently: *"How time flies; another ten days and I have achieved nothing. It doesn't come off. A page now and then is successful, but I can't keep it up, the next day I am powerless"*. Although the burden of writing can be eased by digital writing tools — such as Grammarly, QuillBot or Wordtune — to assist with basic language tasks (such as spelling and grammar checking, paraphrasing, and providing suggestions on style, tone, clarity and coherence), the tools often lack nuance and fall short when more substantive writing assistance is needed. Professional writing services offer advanced editing, rewriting, and even writing from scratch; but they are not accessible to those with limited financial resources and to those who need it most, such as non-native English researchers in economically disadvantaged regions. This exacerbates a communication bottleneck that hampers scientific progress[1,2].

Tools leveraging generative artificial intelligence (generative AI), and particularly generalist large language models (LLMs), are truly useful as writing assistants. LLMs are widely available, versatile, and come across as collaborative, patient and largely non-judgmental[3]. However, using them for effective and responsible writing demands more expertise than the use of specialized digital writing tools, including choosing the appropriate AI systems and crafting effective prompts[4–6]. With this Comment I aim to help researchers to tap into the potential of AI to elevate the quality and efficiency of academic writing and to make the writing process feel less like drudgery and more joyful. In the following sections I introduce a collaborative framework for AI engagement, delineate effective routines and prompting techniques, and discuss major caveats stemming from AI ethics and AI policy. Although written with biomedical and behavioural scientists in mind, the methods I introduce here are applicable to scholarly writing in general.

**A writer–AI collaborative framework**
LLMs power ChatGPT, Gemini, Claude and other popular AI chatbots. They are trained using extensive corpora, with some also including multimodal datasets (visual, audio and textual inputs). They excel at understanding and responding to language-based tasks through user queries or prompts. Although the best way to engage with these tools in science and education is being intensely debated[3,4,7–11], they are set to transform academic research[3,12].

In this section, I outline a collaborative framework for AI engagement in academic writing. The framework addresses (i) the short-term and long-term rationale for engagement along with their underlying mechanisms and potential limitations, (ii) the role of AI throughout the writing process, conceptualized through a two-stage model for writer–AI collaborative writing and, (iii) the nature of AI assistance in writing, through a model of types and levels of writing assistance.

**Rationale.** The importance of engaging with AI can be distilled into short-term and long-term goals. Examples of short-term goals are enhancing writing productivity, quality and creativity for specific projects. AI can make writing faster and better, enabling greater focus on content and ideas — and thereby reducing communication costs and enhancing creativity and quality[9,13–15]. There are two central mechanisms underpinning these gains: cognitive offloading, and imaginative stimulation. Cognitive offloading works similarly to how calculators or statistical programs assist by crunching numbers. By offloading to the AI some of the mental work (from mechanical elements such as spelling and grammar to more nuanced tasks such as word choices, naming and summaries), writers can free up cognitive resources for other tasks. Cognitive offloading improves performance and reduces the likelihood of errors.

Imaginative stimulation works through the synergistic collaboration between writers and AI (not unlike the inventive camaraderie among Renaissance artists, scientists and patrons, which fuelled a period of explosive creativity and discovery). LLMs can present new perspectives, ideas and framing that can spark creativity. For example, in brainstorming sessions, an LLM can serve as a bouncing board, offering counterarguments to challenge prevailing notions. In writing-evaluation sessions, LLMs can provide constructive feedback to help refine the exposition, and may offer interdisciplinary perspectives, challenging the writer to think beyond conventional paradigms. Such interaction, which should be iterative, can broaden the intellectual space for the writer, and promote dynamism that may lead to more well-rounded arguments and to novel insights and solutions.

Long-term goals involve the educational aspects of engaging with LLMs for academic writing[16,17], and are rooted in the age-old educational philosophy of 'learning by doing', which LLMs modernize and maximize by means of two distinctive advantages: adaptive interactivity and instant availability. LLMs offer a personalized and interactive learning experience that caters to the individual style and learning pace of each writer. The adaptability of LLMs is not dissimilar from the one-to-one tutoring model, which has long been recognized for its effectiveness in personalizing education to the learner's needs. By providing real-time feedback to the writer (as in traditional writing workshops), the LLM further helps the writer recognize and correct mistakes, enhances their writing skills, and facilitates a deeper understanding of language.

Instant availability provides an immediate on-demand learning environment, making learning flexible, hands-on and accessible. This echoes the ethos of online open-access educational resources, which democratize learning by making learning materials readily available to all. Still, access to the most powerful LLMs (such as GPT-4) requires a subscription and may not always be available in all regions.

The use of LLMs in academic writing also comes with its drawbacks and limits. First, although LLMs can produce articulate and authoritative-sounding text, their users ought to verify the accuracy of the generated content through fact-checking and cross-referencing[3,11,18]. Responsible use also entails providing appropriate references and adhering to ethical guidelines and disclosure practices[11].

Second, overdependence on LLMs for writing risks could deteriorate writing skills. AI assistance in writing should be used with a critical eye and the intent to learn via active thinking, making decisions about the tool's output, deciding on proper implementation, and noticing and understanding the changes made by the LLM (rather than blindly accepting its output).

Moreover, the quintessential elements of great writing — insight, originality and creativity — cannot yet be fully replicated by LLMs. Even though AI can assist the writing process by facilitating cognitive offloading and imaginative stimulation, and helping summarization, explanation, brainstorming and other tasks[3,14,15], good writing cannot rescue poor scholarship.

**Two-stage collaborative writing.** How can LLMs best help the process of academic writing? The writing process first involves developing the essence of the story, its value proposition, and its structural outline[19–22]. By way of example, the story of this Comment is that academic writing has always been difficult, and that LLMs can elevate its quality and efficiency. This Comment's value proposition lies in providing a timely and much-needed conceptual framework for writer–AI collaborative writing, along with practical techniques to effectively harness AI in academic writing. The structural outline of the piece includes the writer–AI collaborative framework, a table summarizing the levels of writing assistance, and a box with a list of prompts. Therefore, in this AI-inspired stage, the writer solicits imaginative and creative suggestions from the LLM. This interaction expands the writer's conceptual horizons and fuels creative thinking, which can lead to better ideas and structures.

The second stage of the writing process involves the iterative drafting of a manuscript. In this AI-assisted stage, the writer taps into the LLM's linguistic capabilities for assistance, typically by eliciting critical feedback from the LLM, which acts as a second set of eyes to identify potential areas for refinement and clarification. The writer can capitalize on the LLM's linguistic prowess to improve spelling and grammar accuracy, and to enhance word choice, clarity and flow. Incidentally, statistical and drawing software (which can also empower writing) may also leverage AI-based assistance (particularly in problem solving[3] and code writing[9,23]).

Therefore, the LLM can be woven into the entire fabric of the writing process, assisting in both high-level creative tasks (such as brainstorming and critical feedback) and in low-level compositional tasks (such as language polishing and line editing). In this iterative, human-centric dialogue, the writer integrates or discards the LLM's suggestions as deemed fit.

**Types and levels of writing assistance.** A critical point — one that is also central to discussions in AI ethics — is the nature of AI assistance in writing[3,14]. In other words, what types and levels of assistance in writing should AI provide?

First, the assistance can be indirect, where the LLM functions as a writing coach, providing suggestions and feedback without directly altering the text itself. This requires users to decide whether to accept the suggestions and how to implement them. (The coaching role has roots dating back to ancient pedagogical traditions, where tutors guided learning through the Socratic method of questioning and dialogue. Modern writing centres have expanded this coaching capacity, with tutors reviewing and advising on drafts. The AI tool as a writing coach may be considered the digital evolution of these time-honoured educational approaches.)

In direct assistance, the LLM serves instead as a co-writer, contributing original text for the manuscript. Throughout history, creative collisions between authors have fuelled literary and scientific feats (most notoriously, the Goncourt brothers co-writing famed novels, and the cross-continental correspondence between Darwin and Wallace catalysing modern evolutionary theory). By digitally extending the collaborative scope, LLMs can propel this tradition forward.

The degree of assistance can be categorized into five levels. Each successive level exhibits a greater degree of interactive engagement and cognitive contribution from the LLM, as defined in **Table 1** along with example prompts and outputs. In the first level, 'basic editing', the LLM serves as a digital extension of a traditional proof-reader. It ensures language accuracy and lexical variety. In the second level, 'structural editing', the LLM is used as a linguistic artisan that reshapes the text to enhance its clarity and coherence. In the third level, 'creating derivative content', the LLM synthesizes existing content into concise summaries, succinct abstracts or engaging titles. In the fourth level, 'creating new content", the LLM transcends the role of an editor; the tool becomes a collaborative partner that brainstorms and contributes original ideas or alternative perspectives that enrich the narrative and broaden the intellectual horizon. And in the fifth level, 'evaluation or feedback', the LLM acts as a critical reviewer and mentor, offering feedback, analysing strengths and weaknesses, and providing constructive suggestions for refinement and improvement. Across all these levels, assistance can manifest either directly, by providing textual revisions and content that can be used in the manuscript, or indirectly, by offering only suggestions that must be implemented by the writer.

**Routines and techniques**
What effective routines and prompting techniques can best help unlock the potential of AI throughout the writing process? The educational utility of AI can be more effectively realized through indirect assistance (prompt 5 in **Table 1**), whereas shorter-term and project-specific benefits are best achieved through direct assistance (prompts 1–4 in **Table 1**). The type of assistance can be selectively activated through prompting[6], by instructing the LLM to either offer suggestions without altering the text (indirect assistance) or to implement these suggestions by revising the text or adding new content (direct assistance).

The implementation of the various levels of assistance is also outlined in **Table 1**. To do so effectively, one must consider how AI assistance fits organically into the writing routine. Academic writing takes many forms, and is shaped by discipline, genre, message, audience and the writers themselves[24–27]. Writing a brief empirical report on a new phenomenon differs from crafting an elaborate opinion piece on a contentious subject. Some writers draft quickly and revise fiercely, whereas others iterate section by section. Yet all writers must start with a point, however vague, that conveys a message (stage 1), even if they end up deviating from it in a later draft (stage 2). As F. Scott Fitzgerald put it, *"You don't write because you want to say something; you write because you've got something to say."*

Accordingly, an effective routine consists of three iterative processes: outlining the manuscript (broadly defined to also include the message), writing the content, and editing the draft. Each process benefits from various levels of assistance that can be triggered by the prompting techniques provided in **Box 1**. To maximize their utility, the prompts are designed to be specific for each task yet in the context of academic writing.

The optimal degree of assistance depends on the project and on the author's needs. By way of example, for this Comment I had a clear concept and narrative, and hence I did not use LLMs for outlining the manuscript. However, I did extensively use ChatGPT 4 and Claude 2 to edit the text ('Basic editing' in **Box 1**), and occasionally employed completion and continuation techniques when stuck ('Completing or continuing text' in **Box 1**), which greatly enhanced efficiency. When addressing comments from the journal's editor and the reviewers on ethics, I first catalogued all the essential ethics points to be included. I then sought feedback from ChatGPT on this organization (item 1.2 in **Box 1**), which led me to slightly reorder the points. While writing, I merged certain points for tighter organization. I also consulted Perplexity AI with regard to relevant content (I did not use this LLM's output directly, yet it provided helpful assurance).

Regulations of the use of AI in scholarly publishing are in flux, given the abrupt emergence of generative AI. The types and levels of writing assistance by LLMs discussed in this Comment do not violate current guidelines stipulated by most associations on scholarly publishing, publishers and journals. However, caution may be warranted[28]. As of this writing, the *Lancet* group and Elsevier only permit AI to "improve readability and language of the work". Although I expect this stance to change (*Science* originally forbade the use of AI-generated text on January 26, 2023, but abolished this restriction on November 16, 2023), writers should pay close attention to the relevant policies of their target journals.

**Outlining the manuscript.** The outline of a manuscript serves as a roadmap to guide the writing and editing that follows. Although it does vary in elaboration and detail, the outline must encapsulate the central message while addressing these key considerations: what value does the manuscript offer to the audience[25,26]? How should the manuscript be structured to effectively convey the key points through storytelling[20–22]? Although much of this work must come from the writer (by thinking, reading, talking and drafting), LLMs can assist with the content of the outline, and with its actual writing. LLMs can help ensure that all critical points are covered and that the value to the intended audience is clear through brainstorming (assistance level 4) and feedback (assistance level 5). For example, as illustrated by the prompts in **Box 1**, LLMs can help generate ideas for the outline, identify key points, elaborate on an idea or concept, or clarify the value proposition. Also, LLMs can offer critical feedback on key points, the outline itself, the value proposition, and the narrative flow. The actual writing of the outline can be LLM-assisted through rewriting (assistance level 3) and editing (assistance levels 1 and 2) to ensure a logical narrative flow. **Box 1** provides examples of prompts to use when passing a hastily written outline to an LLM for language polishing and for structuring a compelling story arc.

**Writing the content.** Outlines rarely remain fixed; they undergo adjustments during the iterative writing and editing processes. Yet when an outline is ready, the writing process becomes more structured. LLMs can assist in creating both derivative content (based on existing text from the writer or from another source; assistance level 3) and new content (by following the writer's instructions; assistance level 4). For derivative content, LLMs can be used to transform text by rephrasing it to vary the exposition or to avoid plagiarism, and to summarize lengthy text to help the writer better understand the material or to extract key points from it. For new content, LLMs can assist by completing or continuing a piece of text. For example, when a writer is uncertain about which words or phrases to use in a sentence, the LLM can supply multiple options; if the writing is 'stuck', the LLM can 'take over' to provide text that keeps the narrative moving; and if the writer does have specific ideas for continuing, the LLM can also help expand the text to implement those ideas. Moreover, LLMs can assist in the writing of specific parts of a manuscript. For instance, they can generate topic sentences and transition sentences to improve flow. And when the main text of a manuscript is ready, they can craft potential abstracts or titles for it. For all of these types of assistance, examples and instructions can be added to the prompt, to refine the LLM's output (**Box 1**).

**Editing the draft.** Great papers can emerge from messy first drafts but rarely from sloppy rewriting or editing. LLMs can facilitate this editing process by handling basic tasks, supporting more advanced tasks, and acting as an advisor or a peer reviewer (**Box 1**).

Basic editing tasks involve checking spelling and grammar, suggesting synonyms and antonyms, and enhancing vocabulary (**Table 1**). For example, the prompts in **Box 1** can turn an LLM into a digital proof-reader tailored to the needs of the writer, helping with copyediting and improving the clarity and flow of the text, and educating the writer in the process.

Advanced editing tasks include summarizing key points and organizing them logically, extracting key ideas from the text to suggest examples and analogies that enhance its readability, revising text to fit a specific style, analysing the style of the text to emulate it in revisions, and rewriting introductory or concluding sentences or paragraphs to align them with the context of the manuscript and the writer's taste.

When acting as an advisor or a peer reviewer, an LLM can provide thoughtful and constructive feedback by evaluating the writing to identify its strengths and weaknesses, and by offering concrete ways to improve it. It can act as a non-specialist reader to identify jargon in the text and to enhance the accessibility of the writing. It can also serve as a constructive peer reviewer.

**Ethical and policy considerations**

The use of generative AI in academic research has raised pressing ethical questions that are being intensely debated. In academic writing, because writing assistance — from spell checkers and grammar checkers to professional copy editors — has long been widely accepted, it seems justified for AI to be used for the same types of task and need. In particular, for low-level contributions (levels 1 and 2; **Table 1**), the output of LLMs is akin to that of common writing-assistance software.

However, for high-level contributions (levels 3–5), using the output of LLMs verbatim raises ethical considerations and potentially legal challenges[11]. Extensive use of LLMs for academic writing and not properly attributing their assistance risks presenting AI-generated content as original work, thus raising ethical issues of attribution and, possibly, plagiarism. Currently, most publishers forbid listing an AI system as an author[28], yet extensive use challenges the concept of authorship when AI contributes most of the text[29]. It is likely that authorship guidelines and practices will evolve as AI technology rapidly integrates into academia. Guidelines for the ethical use of AI in academic writing will need to consider that restricting the scope of AI use is hardly enforceable. Still, understanding and managing ethical and policy considerations is integral to the responsible and productive use of AI.

Another issue is that, at present, the copyright holder of output generated by LLMs remains legally ambiguous. Microsoft, Google and other providers of generative AI tools offer their users protection against potential legal risks, whereas OpenAI transfers the ownership and rights of its output to users. Yet the fact that by default most LLMs retain the prompts for training purposes raises concerns about copyright (when the prompt includes copyrighted material) and confidentiality or privacy (when prompts include patient data or data from human participants). Users can opt out of data usage for model training by disabling "Chat history & training" in ChatGPT or "AI Data Usage" in Perplexity AI, or by using LLMs designed to respect data privacy (such as ChatGPT Team or ChatGPT Enterprise).

Moreover, because of the stochastic nature of LLMs and their propensity to hallucinate, the output of LLMs cannot always be easily reproduced. The closed-source nature of most leading LLMs further complicates understanding their biases and inaccuracies. Users must be cognizant of these considerations, and actively mitigate potential biases and inaccuracies (for example, by selecting the most suitable LLMs and by verifying their outputs). The recommended pragmatic approach is to keep track of assistance by LLMs throughout the writing project, to transparently report such assistance in the manuscript itself, and to adhere to journal policies on the use of generative AI. In this respect, current policies can be inconsistent[14,28] or lack clarity with respect to methodological transparency and reproducibility[28].

As for methodological transparency, because editing services do not typically require disclosure of use, it seems reasonable to only disclose the use of LLMs when the level of assistance exceeds that of a typical editing service[28]. This point is also recognized by the International Association of Scientific, Technical, and Medical Publishers, which stated in a white paper published on December 5, 2023 that disclosure should be required only when usage goes beyond basic author support (defined as "refining, correcting, formatting, and editing text and documents"; https://www.stm-assoc.org/new-white-paper-launch-generative-ai-in-scholarly-communications).

**Outlook**
Many researchers wish that 'letting the data speak for itself' — rather than wrestling with each word, sentence and paragraph — would suffice. In this Comment, I have outlined a collaborative framework, techniques and caveats for integrating generative LLMs into academic writing. The framework highlights the versatility of LLMs in stimulating ideas and assisting with language tasks throughout the writing process, and the prompting examples in **Box 1** show how to effectively apply different levels of AI assistance, from basic editing to higher-order content generation and critical feedback. The use of LLMs in scientific writing promises to ease communication bottlenecks and thereby to accelerate scientific progress.

Ultimately, the onus is on authors to leverage LLMs prudently to enhance their writing, and to strike a reasonable balance between human creativity and the capabilities and limitations of LLMs. I hope that this Comment provides a conceptual and practical guide to the benefits and challenges of using LLMs in scientific communication.

**Author information**
Zhicheng Lin[1]
[1]University of Science and Technology of China, 96 Jinzhai Road Baohe District, Hefei, Anhui, 230026, China.
e-mail: zhichenglin@gmail.com

**References**
1. Amano, T. *et al. PLoS Biol.* **21**, e3002184 (2023).
2. Lin, Z. & Li, N. *Perspect. Psychol. Sci.* **18**, 358–377 (2023).
3. Lin, Z. *R. Soc. Open Sci.* **10**, 230658 (2023).
4. Birhane, A., Kasirzadeh, A., Leslie, D. & Wachter, S. *Nat. Rev. Phys.* **5**, 277–280(2023).
5. Thirunavukarasu, A. J. *et al. Nat. Med.* **29**, 1930–1940 (2023).
6. Lin, Z. *Nat. Hum. Behav.* doi:10.31234/osf.io/r78fc (in press).
7. Milano, S., McGrane, J. A. & Leonelli, S. *Nat. Mach. Intell.* **5**, 333–334 (2023).
8. White, A. D. *Nat. Rev. Chem.* **7**, 457–458 (2023).
9. Golan, R., Reddy, R., Muthigi, A. & Ramasamy, R. *Nat. Rev. Urol.* **20**, 327–328 (2023).
10. Casal, J. E. & Kessler, M. *Res. Meth. Appl. Linguist.* **2**, 100068 (2023).
11. Lin, Z. *arXiv:2401.15284* (2024).
12. Wang, H. *et al. Nature* **620**, 47–60 (2023).
13. Dergaa, I., Chamari, K., Zmijewski, P. & Ben Saad, H. *Biol. Sport.* **40**, 615–622 (2023).
14. Hwang, S. I. *et al. Korean J. Radiol.* **24**, 952–959 (2023).
15. Bell, S. *BMC Med.* **21**, 334 (2023).
16. Nazari, N., Shabbir, M. S. & Setiawan, R. *Heliyon* **7**, e07014 (2021).
17. Yan, D. *Educ. Inf. Technol.* **28**, 13943–13967 (2023).
18. Semrl, N. *et al. Hum. Reprod.* **38**, 2281–2288 (2023).
19. Chamba, N., Knapen, J. H. & Black, D. *Nat. Astron.* **6**, 1015–1020 (2022).
20. *Nat. Biomed. Eng.* **2**, 53 (2018).
21. Croxson, P. L., Neeley, L. & Schiller, D. *Nat. Hum. Behav.* **5**, 1466–1468 (2021).
22. Luna, R. E. *Nat. Rev. Mol. Cell Biol.* **21**, 653–654 (2020).
23. Merow, C., Serra-Diaz, J. M., Enquist, B. J. & Wilson, A. M. *Nat. Ecol. Evol.* **7**, 960–962 (2023).
24. Yurkewicz, K. *Nat. Rev. Mater.* **7**, 673–674 (2022).
25. King, A. A. *J. Manag. Sci. Rep.*, doi:10.1177/27550311231187068 (2023).
26. Patriotta, G. *J. Manag. Stud.* **54**, 747–759 (2017).
27. Gernsbacher, M. A. *Adv. Meth. Pract. Psych.* **1**, 403–414 (2018).
28. Lin, Z. *Trends Cogn. Sci.* **82**, 85–88 (2024).
29. Lin, Z. *PsyArXiv*, doi:doi.org/10.31234/osf.io/s6h58 (2023).

**Acknowledgments**
The writing of this Comment was supported by the National Key R&D Program of China STI2030 Major Projects (2021ZD0204200), the National Natural Science Foundation of China (32071045), and the Shenzhen Fundamental Research Program (JCYJ20210324134603010). The author used GPT-4 (https://chat.openai.com) and Claude (https://claude.ai) alongside prompts from Box1 to help write earlier versions of the text and to edit it. The text was then developmentally edited by the journal's Chief Editor with basic-editing and structural-editing assistance from Claude, and checked by the author.

**Competing interests**
The author declares no competing interests.

**Additional information**

**Publisher's note** Springer Nature remains neutral with regard to jurisdictional claims in published maps and institutional affiliations.

**Peer review information** *Nature Biomedical Engineering* thanks Serge Horbach and the other, anonymous, reviewer(s) for their contribution to the peer review of this work.

**Box 1 | Example prompts to invoke the assistance of LLMs in academic writing.** The prompts are meant to guide the LLM by providing it with relevant context, examples or explicit instructions and even emotional cues, and to constrain the LLM's output by instructing it to explore multiple options or to take on certain roles.

### 1. Outlining

#### 1.1 Brainstorming

"Generate potential sections and sub-sections for a manuscript exploring {title or topic}."

"Identify key points to be discussed under the topic of {topic}."

"Expand on the idea of {idea}, detailing {aspects of discussion}."

"What value could the exploration of {topic} bring to the audience of {audience}? Suggest different angles of value proposition for the manuscript."

#### 1.2 Evaluation and feedback

"Evaluate the key points listed under the sections of {sections} in the provided outline for relevance, depth, and alignment with the manuscript's central message of {message}. Outline: {outline}."

"Review the provided outline for the manuscript on {title or topic} and suggest improvements for clarity, coherence, and completeness. Outline: {outline}."

"Assess the value proposition presented in the outline for the manuscript on {title or topic}. Suggest any additional angles or refinements to better appeal to the intended audience of {audience}. Outline: {outline}."

"Analyse the narrative flow of the provided outline for the manuscript on {title or topic}. Provide suggestions for enhancing the storytelling aspect to effectively convey the key points and value proposition. Outline: {outline}."

#### 1.3 Rewriting and editing

"Edit the jotted outline for the manuscript on {title or topic}. Rewrite to make it concise. Outline: {jotted outlines}."

"Propose a narrative structure that effectively communicates the key points and value proposition of {title or topic}, ensuring a coherent, logical flow from introduction to conclusion. Key points: {key points}. Value proposition: {value proposition}."

### 2. Writing content

#### 2.1. Transforming text

"Act as a top editor in top journals. I will provide you with text. Your task is to paraphrase the text. Show 3 versions. Confirm with OK."

"Act as a top editor in top journals. Summarize the text provided. The summary {requirements; e.g., needs to be in about 100 words and cover all the main points}. Text: {text}."

#### 2.2. Completing or continuing text

"Act as a top editor in top journals. I will provide you with text that is missing some parts, marked with {...}. Your task is to fill in the {…} parts. Show 5 variations of completion. Include only the filled-in text in the output format. Confirm understanding with OK."

"Prompt 1: Act as a top editor in top journals. I will provide you with text. Your task is to continue the text based on the instruction I provide. Show 3 versions of continuation. Confirm with OK.

Prompt 2: Instruction: Continue from {end of text}. Explain {what to explain}. Show {points to make}, drawing parallels or analogies where you see fit. End by {what to end}. Text: {text}."

### 2.3. Writing parts

"Act as a top editor in top journals. Provide three alternative topic sentences for each of the following paragraphs. Ensure {requirements; e.g., that they encapsulate the main idea of each paragraph and align with the overall theme of the manuscript}."

"Create three alternative transition sentences between the following two paragraphs:

Paragraph 1: {paragraph}

Paragraph 2: {paragraph}"

"Act as a top researcher writing for top journals. Write three alternative {abstracts, titles, etc.} for the attached manuscript to be submitted to [28]. The {abstract, title, etc.} needs to {requirements}. Two examples {abstract, title, etc.} are given below.

Example 1: {example}

Example 2: {example}"

## 3. Editing the draft

### 3.1. Basic editing

"Act as an expert editor for top scientific journals (Nature, Science) to improve the clarity and flow of the writing. Take a deep breath—this is very important for my career! I will give you text later, and your job is to offer three revisions with a brief explanation of the changes based on the following instructions.

The first version is a copyedited version, with the following rules:

– only make needed changes (e.g., if it is just a matter of subjective choice/preferences, don't make the changes)

– when the text provided is long, copyedit through each paragraph (that is, don't condense the long text unless necessary)

The second and third versions are the ones you revise to maximize clarity and flow by following the key recommendations suggested by the following books, but make the two versions distinct:

1) "The Elements of Style" by William Strunk Jr. and E.B. White

2) "Style: Lessons in Clarity and Grace", by Joseph M. Williams and Joseph Bizup

3) "On Writing Well" by William Zinsser

4) "The Sense of Structure: Writing from the Reader's Perspective" by George Gopen

Your output consists of the three versions of texts and a brief explanation of the changes for each version (omit the text provided to you). I will tip $200 when you provide great responses. Confirm by replying with OK."

"Compare:
{text to compare, such as: 1) But they fall short in more substantive tasks. 2) But they fall short of more substantive tasks.}"

### 3.2. Advanced editing

"Act as an expert editor for top scientific journals. First, summarize the key point of each paragraph. Then, explain the logic of how these points are connected. Finally, provide suggestions on how to organize these points in the most logical way."

"Act as a skilled editor. Help researchers augment their manuscript with examples and analogies to enhance its accessibility and ground ideas in real-world contexts. Your task is to examine the manuscript to extract three central ideas, and then suggest three examples and analogies for each idea."

"Revise the text below, using the style of {style}. Text: {text}."

"Evaluate the following writing style: {text}. Revise the following text based on the style above. Text: {text}"

"Considering the main aim of the manuscript is {aim}, rewrite the following introduction paragraph to better align with this aim and highlight the significance of the study, while ensuring clarity and reader engagement: {paragraph}"

"Offer five alternatives that revise the {first or last} paragraph to make it an {artful} {beginning or end}."

"Considering the broader implications of the research are {implications}, rewrite the following conclusion paragraph to better reflect these implications and emphasize the key findings and the value they add to the field, while ensuring a strong, clear, and concise closure to the manuscript: {paragraph}"

### 3.3. Evaluation and feedback

"Act as an expert editor for top scientific journals. Evaluate the following text. For each weakness provide examples to help strengthen it. Confirm with OK."

"Act as a supportive professor aiding researchers from disparate fields in refining their manuscript. Identify technical terms or jargon, ensuring the text is comprehensible, at least conceptually, to a non-expert like yourself. Clearly and succinctly pinpoint any confusing aspects or unclear points that could perplex others."

"Act as a meticulous reviewer for prestigious journals. Compose a thorough peer review for the attached manuscript, using the outline below:

– Novelty and importance

– Reasons favouring acceptance

– Grounds for rejection: List major reasons, elaborating on each with sub-points to provide detailed justification. Strive for specificity.

– Recommendations for enhancement: Offer several key recommendations, articulating them with precision.

Maintain a thoughtful and constructive approach throughout."

**Table 1 | Five levels of writing assistance by LLMs, with example prompts.** The type of assistance — direct (levels 1–4) or indirect (level 5) — can be invoked by prompting accordingly. For example, for levels 1–4, adding the prompt "Do not revise or rewrite the text, but only offer suggestions for the user to implement" results in indirect assistance; conversely, for level 5, the prompt "Revise the text to implement the suggestions provided" leads to direct assistance.

| Level | Prompt |
| --- | --- |
| **1. Basic editing**, such as checking spelling and grammar, or suggesting synonyms. | "Check the spelling and grammar in this paragraph, and suggest synonyms for any repetitive words." |
| **2. Structural editing**, such as paraphrasing, translating, or improving the structure of the text, or its flow or coherence. | "Paraphrase this lengthy sentence to improve its clarity and flow, and translate it to French." |
| **3. Creating derivative content**, such as summarizing, creating titles and abstracts, rewriting or generating analogies. | "Summarize this document and create a short, catchy title for a journal submission." |
| **4. Creating new content**, such as completing, continuing or expanding text, or brainstorming ideas. | "Continue the text to explain the key question being addressed. Show why it is important, drawing parallels or analogies where you see fit." |
| **5. Evaluation or feedback**, such as assessing the quality of the writing or finding weaknesses in it. | "Review this introduction and highlight any logical gaps or areas that need further development." |

These five levels exemplify the utility of the LLMs that are publicly available as of January 2024.